\documentclass[aps,showpacs,prc,twocolumn]{revtex4}
\usepackage{graphicx}
\usepackage{amssymb}
\begin{document}

\title{Reaction dynamics of weakly-bound nuclei at near-barrier energies: impact of incomplete fusion on the angular distribution of direct alpha-production}

\author{Alexis Diaz-Torres}
%\ead{alexis.diaz-torres@anu.edu.au}
%\author{J.A. Tostevin}
\affiliation{Department of Physics, Faculty of Engineering and
Physical Sciences, University of Surrey, Guildford, Surrey GU2 7XH,
United Kingdom}

\begin{abstract}
The classical trajectory model with stochastic breakup for nuclear collision dynamics of weakly-bound nuclei is further developed. It allows a quantitative study of the importance of incomplete fusion dynamics in the angular distribution of direct alpha-production. Model calculations indicate that the incomplete fusion contribution diminishes with decreasing energy towards the Coulomb barrier, notably separating in angles from the contribution of no-capture breakup events. This should facilitate the experimental disentanglement of these competing reaction processes. 
\end{abstract}
\maketitle

\emph{Introduction}.
Nuclear physics research has entered a new era with developments of radioactive nuclear beam 
facilities, where nuclear reactions are the primary probe of the new physics, such as novel structural changes. In those facilities, the low-energy nuclear reactions research is highly focused on understanding astrophysically-important reaction rates involving exotic nuclei. These are often weakly-bound with a few-body, cluster structure that can easily be dissociated in their interaction with other nuclei. Understanding the breakup mechanism and its impact on nuclear reaction dynamics is essential. A major consequence of breakup is that a rich scenario of reaction pathways arises, such as events where (i) not all the resulting breakup fragments might be captured by the target, termed incomplete fusion ({\sc icf}), (ii) the entire projectile is captured by the target, called complete fusion ({\sc cf}), and (iii) none of the breakup fragments are captured, termed no-capture breakup ({\sc ncbu}). 

Since the availability of intense exotic beams is still limited, extensive experimental research has recently been carried out exploiting intense beams of stable weakly-bound nuclei, such as $^{6,7}$Li and $^{9}$Be \cite{Nanda1,Nanda2,Beck0}. Understanding the effect of their breakup on near-barrier fusion has been a key aspect of these investigations \cite{Canto1}. These have definitively demonstrated that breakup suppresses the above-barrier fusion cross sections. Most recently, experimental activities are focused on disentangling breakup and competing reaction mechanisms from inclusive and exclusive coincidence measurements \cite{Signorini1,Shrivastava,Beck1,Santra1,Souza1}. A challenge is to obtain a complete quantitative understanding of the breakup mechanism and its relationship with near-barrier fusion. This research is guided by complete sub-barrier breakup measurements \cite{Nanda3}.  

Theoretical works have addressed the low-energy reaction dynamics of weakly-bound nuclei using quantum mechanical, classical and mixed quantum-classical approaches \cite{Jeff1,CDCC1,TCSM1,CDCC2,Yabana,Hagino,CDCC3}. Among these, the continuum-discretised coupled channels ({\sc cdcc}) framework has been very successful \cite{Jeff1,CDCC2,CDCC3}. However, existing quantum models have limitations \cite{Ian1}, as they cannot calculate integrated {\sc icf} and {\sc cf} cross sections unambiguously. Neither, after the formation of {\sc icf} products, can they follow the evolution of surviving breakup fragment(s) since {\sc icf} results in depletion of the total few-body wave function. 

These difficulties are overcomed by the three-dimensional classical dynamical reaction model suggested in Ref. \cite{Alexis1}. It allows a quantitative description of {\sc icf} and {\sc cf} of weakly-bound (two-body) projectiles, relating the sub-barrier {\sc ncbu} to the above-barrier {\sc cf} suppression.

I have further developed this approach. The key new aspect is the time propagation of the surviving breakup fragment and the {\sc icf} product, allowing the description of their asymptotic angular distribution and dynamical variables. This development should be very useful in (i) current experimental activities aimed at disentangling competing reaction mechanisms from asymptotic observables such as alpha-production yields \cite{Souza1}, and (ii) applications to $\gamma$ ray spectroscopy \cite{Dracoulis1,Gasques1}. The approach is quantitatively illustrated with a simplified test case. The methodology will be explained first. Afterwards, results are presented and discussed, and finally a summary is given.   

\emph{Methodology}. Details of the approach are given in Ref. \cite{Alexis1}, and its implementation in the {\sc platypus} code is described in Ref. \cite{Alexis2}. Only the main features of the model are highlighted here: 
\begin{description}
\item \textnormal{(i)} The target $T$ is initially at rest in the origin of the laboratory frame, whilst the weakly-bound (two-body) projectile $P$ approaches the target (along the z-axis) with incident energy $E_0$ and orbital angular momentum $L_0$. For each $L_0$ (chosen to be an integer number of $\hbar$) an \emph{ensemble} of $N$ incident projectiles is considered. Including the $P-T$ mutual Coulomb and nuclear forces, classical equations of motion determine an orbit with a definite distance of closest approach $R_{min}(E_0,L_0)$.

\item \textnormal{(ii)} The complexity of the projectile dissociation is empirically encoded in a density of (local breakup) probability ${\cal P}^L_{BU} (R)$, a function of the projectile-target separation $R$, such that ${\cal P}^L_{BU}(R) dR$ is the probability of breakup in the interval $R$ to $R+dR$ (see Appendix A in \cite{Alexis2}). A key feature is that for a given projectile-target combination, both experimental measurements \cite{Hinde1} and CDCC calculations \cite{Alexis1} indicate that the integral
of this breakup probability density along a given classical orbit is an
exponential function of its distance of closest approach,
$R_{min}(E_0,L_0)$: 
\begin{eqnarray}
{P}_{BU}(R_{min})&=&2\int_{R_{min}}^\infty {\mathcal P}^L_{BU}(R) dR \nonumber\\ 
                 &=&A\,\exp (-\alpha R_{min}). \label{one}
\end{eqnarray}
Consequently, ${\cal P}^L_{BU} (R)$ has the same exponential form, ${\mathcal P}^L_{BU}(R) \propto \exp (-\alpha R)$. It is sampled to determine the position of breakup in the orbit discussed in (i). In this position, the projectile is instantaneously broken up into fragments F1 and F2. These interact with $T$, and with each other, through real central two-body potentials having Coulomb barriers
$V_{B}^{ij}$ at separations $R_{B}^{ij}$, $i,j=1,2,T,\ i\neq j$. 
The instantaneous dynamical variables of the excited projectile at
breakup, namely its total internal energy $\varepsilon_{12}$, its angular
momentum $\vec{\ell}_{12}$ and the separation of the fragments
$\vec{d}_{12}$ are all Monte Carlo sampled \cite{Alexis1}. 
Having fixed the position and dynamical variables of the excited projectile 
fragments at the moment of breakup, the instantaneous velocity of the particles
F1, F2 and $T$ is determined by conservation of energy, linear
momentum and angular momentum in the overall center-of-mass frame 
(see Appendix B in \cite{Alexis2}). 
These breakup initial conditions are transformed to the laboratory frame where the
three bodies are propagated in time. The calculated trajectories of F1, F2 and $T$ determine the number of {\sc icf}, {\sc cf} and {\sc ncbu} events, fragment F$j$ being assumed to be captured if the classical trajectories take it within the fragment-target barrier radius $R_{B}^{jT}$.

\item \textnormal{(iii)} From the $N$ breakup events sampled for each projectile 
angular momentum $L_0$, the numbers of events $N_i$ in which $i=$ 0
({\sc ncbu}), 1 ({\sc icf}), or 2 ({\sc cf}) fragments are captured
determine the relative yields $\widetilde{P_i}=N_i/N$ of these three
reaction processes after breakup, with $\widetilde{P_{0}} +
\widetilde{P_{1}} + \widetilde{P_{2}} = 1$. The absolute probabilities
$P_i(E_0,L_0)$ of these processes are expressed in terms of the 
relative yields and the integrated breakup probability over the whole 
trajectory $P_{BU}(R_{min}$):
\begin{eqnarray}
P_{0}(E_0,L_0) &=& P_{BU}(R_{min})\,\widetilde{P_{0}},\label{eq5} \\
P_{1}(E_0,L_0) &=& P_{BU}(R_{min})\,\widetilde{P_{1}},\label{eq3} \\
P_{2}(E_0,L_0) &=& [1-P_{BU}(R_{min})]\,H(L_{cr}-L_0) \nonumber\\
               &+& P_{BU}(R_{min})\, \widetilde{P_{2}},
\label{eq1}
\end{eqnarray}
where $H(x)$ is the Heaviside step function and $L_{cr}$ is the critical partial wave for projectile fusion. The cross sections are calculated using
\begin{equation}
\sigma_i (E_0)= \pi \lambda ^2 \sum_{L_0} (2L_0 + 1) P_i(E_0,L_0),
\label{eq2}
\end{equation}
where $\lambda ^2=\hbar^2/[2m_P E_0]$ and $m_P$ is the projectile mass.      
\end{description} 

Beside the absolute cross sections (\ref{eq2}), asymptotic observables, 
such as the angle, kinetic energy and
relative energy distributions of the fragments from {\sc ncbu} 
events, are calculated by tracking their trajectories to a large 
distance from the target. 
%which was 200 fm in the calculations given below.

Crucially, for the {\sc icf} events, the time propagation of the {\sc icf} product and the surviving breakup fragment is now incorporated into this picture. The captured fragment reaches the target radius forming the {\sc icf} product, whilst the other fragment flies away. At this moment, the three-body propagation turns into a two-body propagation, with definite interaction potentials and initial conditions. These are given by the position and velocity of the three particles, at the moment when the {\sc icf} product is formed.

This approach is here applied to the test reaction $^8$Be + $^{208}$Pb. This is because a crucial model input is the local projectile breakup probability ${\cal P}^L_{BU} (R)$, whose form deduced from ${P}_{BU}(R_{min})$ in 
(\ref{one}) is known thus far only for the reaction $^9$Be + $^{208}$Pb from sub-barrier breakup measurements \cite{Hinde1}. Very slightly changing the empirical 
${\cal P}^L_{BU} (R)$ from the $^9$Be + $^{208}$Pb experiment, the classical model {\sc cf} and {\sc icf} excitation functions for $^8$Be + $^{208}$Pb very well agree with the experimental data for 
$^9$Be + $^{208}$Pb at above-barrier energies (see Figs. 1 and 3 in 
\cite{Alexis1}). Thus, the present outcomes for the angular distribution of direct alpha-production are expected to represent those of the $^9$Be + $^{208}$Pb reaction. The optimal breakup function ${P}_{BU}(R_{min})$ has parameters $A=5.98\times 10^3$ and $\alpha=0.85$ fm$^{-1}$ [see Eq.\ (\ref{one})]. The nuclear interaction between the alpha particle and the {\sc icf} product $^{212}$Po is the Woods-Saxon (WS) potential well ($V$, $r$, $a$) $\equiv$
(33.98 MeV, 1.48 fm, 0.63 fm) deduced from the global Broglia-Winther parametrization \cite{Broglia1}. (Please note that in the potential the radius parameter is multiplied by $A_T ^{1/3}$.) The rest of the model parameters are the same as in Ref. \cite{Alexis1}.

\begin{figure}
\begin{center}
%\begin{tabular}{cc}
\includegraphics[width=0.40\textwidth,angle=0]{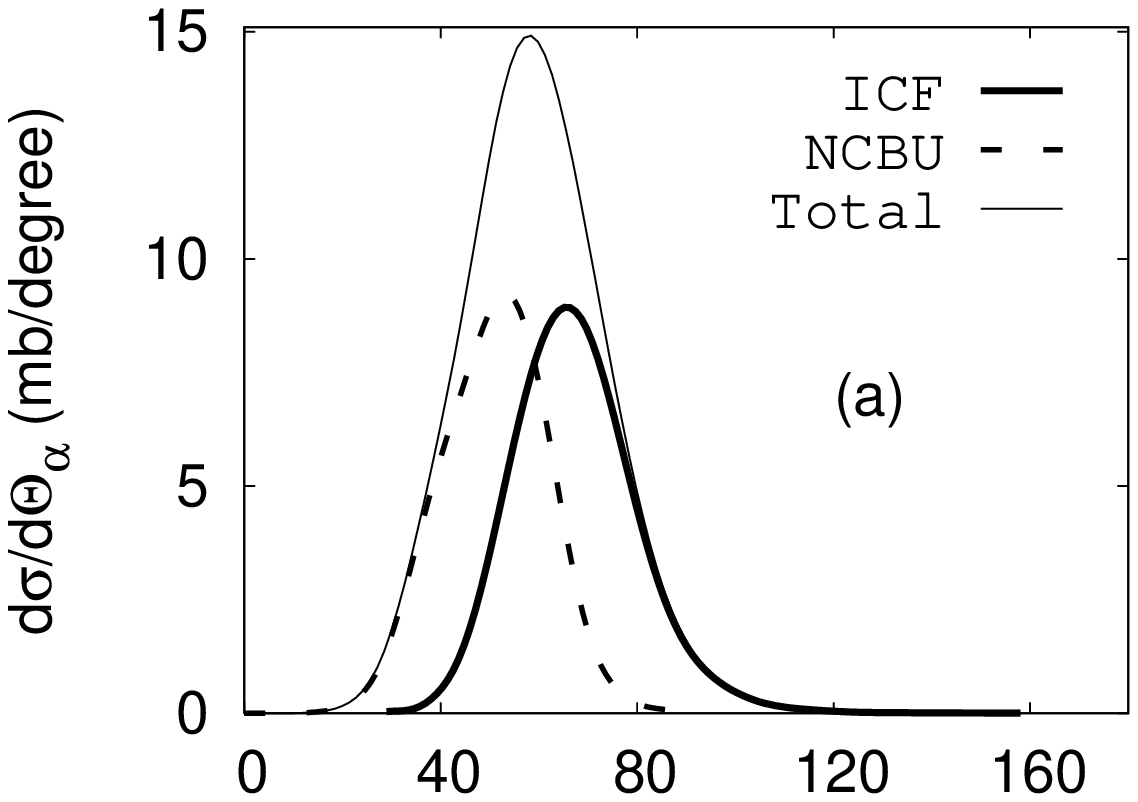} \\ 
\includegraphics[width=0.40\textwidth,angle=0]{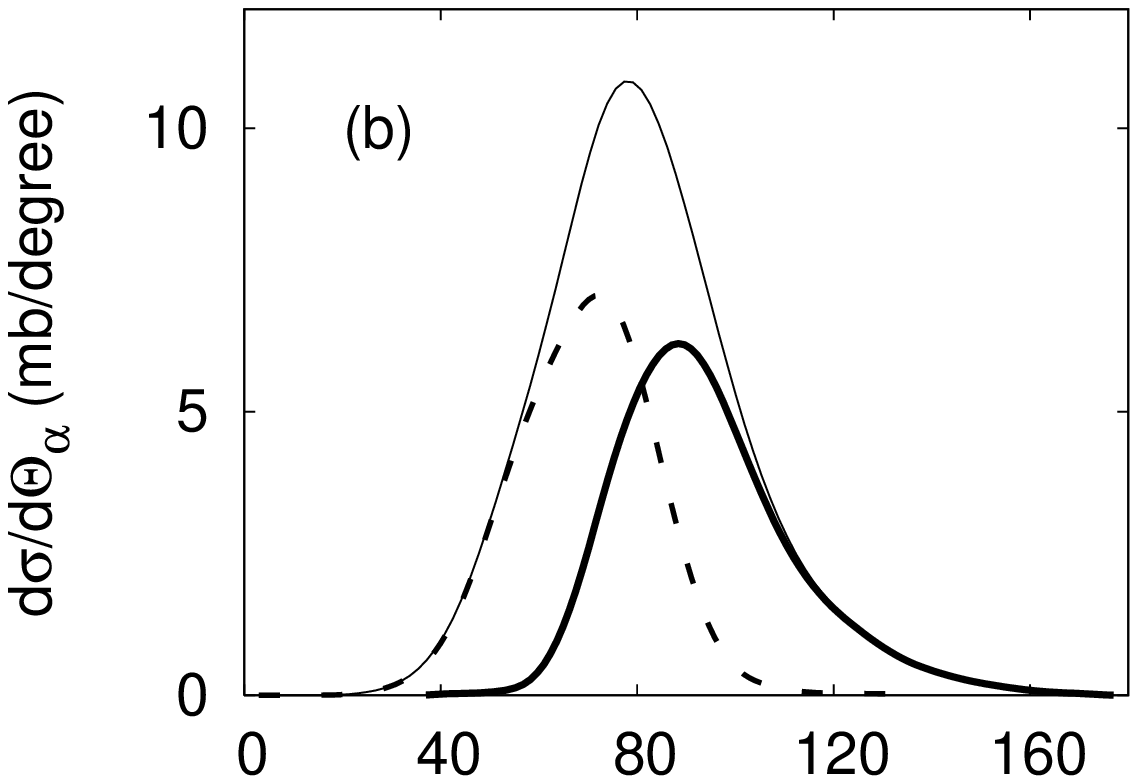} \\
\includegraphics[width=0.40\textwidth,angle=0]{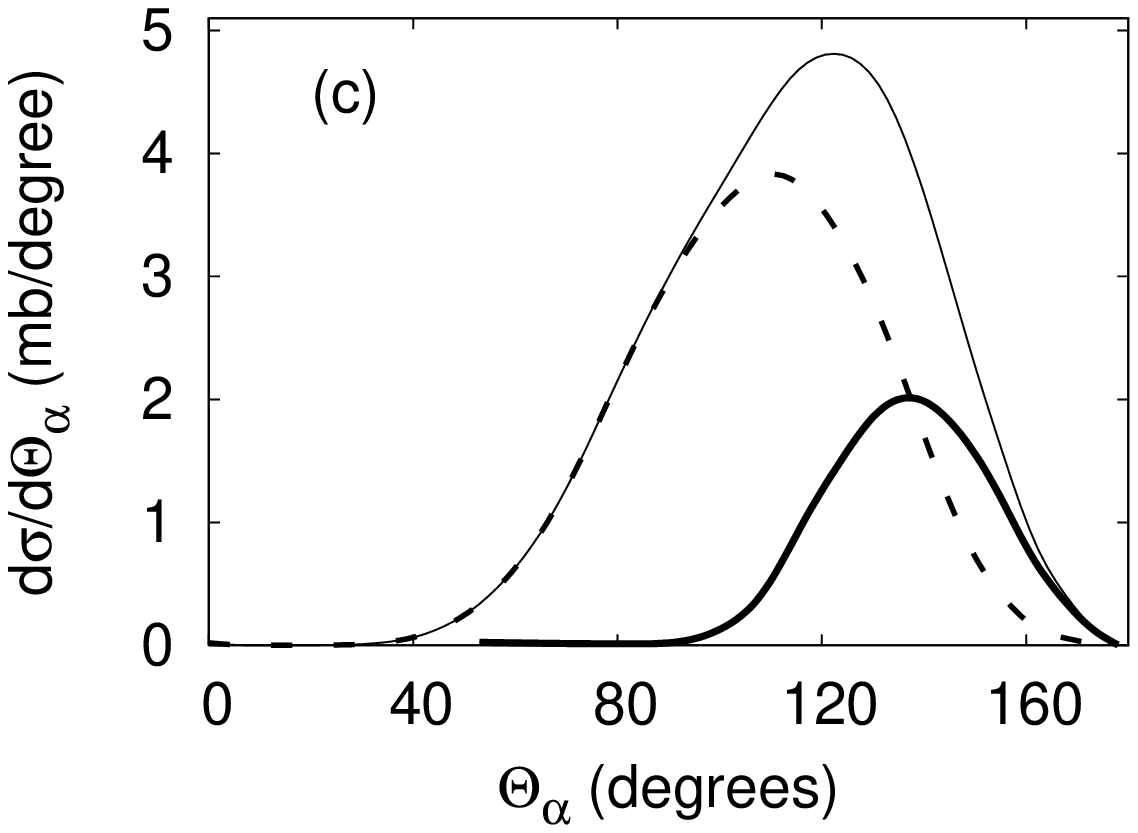}
%\end{tabular}
\caption{Angular distribution of direct alpha-production 
for $^8$Be + $^{208}$Pb for three laboratory energies $E_0$: 
(a) 65 MeV, (b) 55 MeV, and (c) 45 MeV. 
With decreasing energy towards the s-wave Coulomb barrier, 
the {\sc ncbu} events dominate, separating its 
centroid substantially from that of the {\sc icf} events. 
The total alpha-production distribution becomes asymmetric.} 
\label{Figure1}
\end{center}
\end{figure}

\begin{figure}
\begin{center}
%\begin{tabular}{cc}
\includegraphics[width=0.50\textwidth,angle=0]{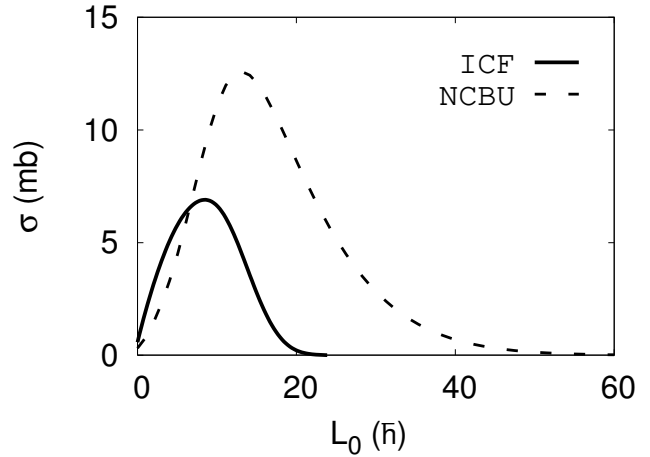}
%\end{tabular}
\caption{{\sc icf} and {\sc ncbu} cross sections as a function of 
the relative angular momenta $L_0$ for $^8$Be + $^{208}$Pb 
at $E_0 = 45$ MeV. The contribution of high-partial waves 
shifts the {\sc ncbu} distributions in Fig. 
\ref{Figure1} to smaller angles, with respect to the 
{\sc icf} distributions.} 
\label{Figure2}
\end{center}
\end{figure}

\emph{Results and discussion}. Figure \ref{Figure1} shows the angular distribution of 
direct alpha-production for three incident, laboratory energies near the $P-T$ s-wave Coulomb barrier ($39.9$ MeV), namely (a) $E_0 = 65$, (b) $55$, and (c) $45$ MeV. The contribution of the {\sc icf} and {\sc ncbu} events is represented by thick solid and thick dashed lines, respectively. Their sum is represented by the thin solid line. It has a symmetric shape at energies well-above the barrier [panels (a) and (b)], adopting an asymmetric form as the incident energy decreases [panel (c)]. While the contribution of the {\sc icf} and {\sc ncbu} events appears to be similar at well-above barrier energies, the {\sc ncbu} contribution gradually dominates with decreasing energy towards the barrier. Here, its centroid significantly separates from the centroid of the {\sc icf} contribution. Both centroids shift to higher angles as the incident energy decreases, due to the reduction of relative partial waves affecting these reaction processes. However, the {\sc ncbu} centroid always remains lower than the {\sc icf} centroid, as higher partial waves contribute to the {\sc ncbu} process (see Fig. \ref{Figure2}). Fig. \ref{Figure2} presents the incident angular momentum distribution of the {\sc icf} (solid line) and {\sc ncbu} (dashed line) processes at a laboratory energy of $E_0 = 45$ MeV. For completeness, the absolute {\sc icf}, {\sc ncbu} and {\sc cf} cross sections are given in Table \ref{Table} for the laboratory energies studied.

\begin{table} \caption{The absolute {\sc icf}, {\sc ncbu} and {\sc cf} cross sections for the laboratory energies studied.} 
\label{Table}
\medskip
\begin{center}
\begin{tabular}{c|ccc}
\hline\hline 
%&\multicolumn{2}{c}{State $|X \rangle $} &
%\multicolumn{2}{|c}{States $|X \rangle $ and $|Y\rangle$}
%\\
%\hline
$E_0$ (MeV) &~~~$\sigma_{ICF}$ (mb)~~&~~$\sigma_{NCBU}$ (mb)~~ &~~ 
$\sigma_{CF}$ (mb)~\\
\hline
45 &84.7&238.84&149.2\\
55 &252.3&255.1&624.6\\
65 &273.6&259.3&981.5\\
\hline\hline \end{tabular}
\end{center}
\end{table}

\emph{Summary}. The classical trajectory model with stochastic breakup has been extended to the calculation of asymptotic observables associated with the {\sc icf} process. This development should be very useful for separating the contribution of competing reaction mechanisms of weakly-bound nuclei from inclusive and exclusive measurements, such as the total alpha-production yield. This may also be affected by other direct processes, such as transfer \cite{Alexis3}, which are not included in the model yet. The inclusion of transfer as well as the development of a unified quantum description remain as great theoretical challenges. The former may be carried out exploiting the concept of transfer function \cite{Oertzen}, and the latter through a time-dependent density-matrix approach incorporating the concept of quantum decoherence \cite{Ian1,Alexis4}. Nevertheless, for the first time, the impact of {\sc icf} dynamics on the angular distribution of direct alpha-production at near-barrier energies is quantitatively presented. The {\sc icf} contribution diminishes with decreasing energy towards the barrier, making the direct, total alpha-production distribution asymmetric. The {\sc icf} and {\sc ncbu} contributions are clearly separated in angles at energies close to the barrier, facilitating the experimental disentanglement of these competing reaction processes. 
 
\begin{acknowledgments}
Support from the UK Science and Technology Facilities Council (STFC) Grant No. ST/F012012/1 is
acknowledged.
\end{acknowledgments}

\end{document}